\documentclass{IEEEtran}

\usepackage{cuted}

\usepackage{lscape}

\usepackage{cite}
\usepackage{graphicx}
\usepackage[cmex10]{amsmath}
\usepackage{amssymb}
\usepackage{booktabs}
\usepackage{threeparttable}
\usepackage{algorithm}
\usepackage{algpseudocode}
\usepackage{amsfonts}
\usepackage{xcolor}
\usepackage{color}

\usepackage{tabularx} %

\usepackage{caption}

\usepackage{enumitem}
\usepackage{framed}

\usepackage{dblfloatfix} %

\usepackage{soul}

\usepackage{cancel} %

\usepackage{verbatim}%

\usepackage{subfigure}

\usepackage{makecell}

\usepackage{empheq} %

\usepackage{underoverlap} %

\usepackage{hyperref}
\hypersetup{pdftex,colorlinks=true,allcolors=black}

\usepackage{calc}

\usepackage{textcomp}
\def\BibTeX{{\rm B\kern-.05em{\sc i\kern-.025em b}\kern-.08em
		T\kern-.1667em\lower.7ex\hbox{E}\kern-.125emX}}

\newcommand{\dfclk}{\Delta f_{\mathrm{CLK}}} %
\newcommand{\dfs}{\Delta f_{\mathrm{s}}} %
\newcommand{\dts}{\Delta T_{\mathrm{s}}} %
\newcommand{\domega}{\Delta \omega} %
\newcommand{\md}{\mathsf{d}} %
\newcommand{\dtih}{\Delta t_{ih}} %
\newcommand{\dtzero}{\Delta t_{0}} %
\newcommand{\oih}{\omega_{ihm}} %
\newcommand{\varpiIhm}{\varpi_{ihm}} %

\newcommand{\ms}{\mathrm{s}} %
\newcommand{\mt}{\mathrm{t}} %
\newcommand{\mr}{\mathrm{r}} %
\newcommand{\mpp}{\mathrm{p}} %
\newcommand{\oihm}{\omega_{ihm}} %
\newcommand{\myModulo}[2]{\left\langle#1\right\rangle_{#2}} %

\newtheorem{lemma}{\bf  Lemma}

\newcommand{\mj}{\mathsf{j}}

\begin{document}

	\title{\huge Practical Frequency-Hopping MIMO Joint Radar Communications: Design and Experiment}

	\author{Jiangtao Liu, 
		Kai Wu, Tao Su and J. Andrew Zhang 
		
	\thanks{Jiangtao Liu and Tao Su are with National Laboratory of Radar Signal Processing, Xidian University, Xi'an Shaanxi 710071, China (e-mail: jiangtaoliu@xidian.edu.cn;
				sutao@xidian.edu.cn)}

	\thanks{Kai Wu and J. Andrew Zhang are with the Global Big Data Technologies Centre, University of Technology Sydney, NSW 2007, Australia (e-mail: kai.wu@uts.edu.au; andrew.zhang@uts.edu.au)}
	}

	\maketitle

	\begin{abstract}	
		Joint radar and communications (JRC) can realize two radio frequency (RF) functions using one set of resources, greatly saving hardware, energy and spectrum for wireless systems needing both functions. 	
		Frequency-hopping (FH) MIMO radar is a popular candidate for JRC, as the achieved communication symbol rate can greatly exceed radar pulse repetition frequency. However, practical transceiver imperfections can fail many existing theoretical designs. In this work, we unveil for the first time the non-trivial impact of hardware imperfections on FH-MIMO JRC and analytically model the impact. We also design new waveforms and, accordingly, develop a low-complexity algorithm to jointly estimate the hardware imperfections of unsynchronized receiver. Moreover, employing low-cost software-defined radios and commercial off-the-shelf (COTS) products, we build the first FH-MIMO JRC experiment platform with radar and communications simultaneously validated over the air. 
		Corroborated by simulation and experiment results, the proposed designs achieves high performances for both radar and communications. 
		
	\end{abstract}
	

\section{Introduction}

The proliferation of wireless systems has caused severe spectrum congestion and scarcity worldwide. 
To alleviate the issue, joint radar and communications (JRC) has been identified as a promising solution \cite{IoT_cui2021integrating,FanLiu_Overview2019}. 
By sharing waveform, spectrum frequency, hardware and signal processing modules, JRC can substantially improve cost, energy and spectral efficiency of wireless systems that require both sensing and communications functions \cite{Kai_pmn}. One of the major JRC designs is radar-centric by integrating data communications into existing radar platforms \cite{Kai_overview_fhMIMO_dfrc2021}. Such design is also referred as dual-function radar-communication (DFRC) in the open literature \cite{DFRC_SP_Mag2019Amin_Aboutanios}.

Initial DFRC works, e.g., \cite{1211013,4268440,barrenechea2007fmcw}, employ the linear frequency-modulated (LFM) signal-based pulsed radars given their wide applicability in radar community. 
In general, these works \cite{1211013,4268440,barrenechea2007fmcw} employ the frequency modulation rate, e.g., positive and negative, to convey one communication symbol per radar repetition time (PRT). 
To increase the communication symbol rate, more recent DFRC designs lean toward using MIMO radars due to their rich degree of freedom (DoF) in waveform design. For example, 
beam patterns of a MIMO radar is optimized to exploit sidelobes to conduct communication modulations, e.g., phase shift keying (PSK) and amplitude shift keying \cite{DFRC_SidelobeControl2016TSP,DFRC_SparseArray2019TAES_XianrongWang}. 
The MIMO radar waveform has also been optimized to conduct non-conventional modulations, e.g., code shift keying \cite{DFRC_CSK2018DSP} and waveform shuffling \cite{DFRC_waveformShuffling2018DSP}. 
Despite that more information bits can be carried per symbol (compared with initial LFM-based DFRC designs), 
these works \cite{DFRC_SidelobeControl2016TSP,DFRC_SparseArray2019TAES_XianrongWang,DFRC_waveformShuffling2018DSP,DFRC_CSK2018DSP} still embed one information symbol over one or several radar pulses. Thus, their achieved symbol rate is still limited by the radar pulse repetition frequency (PRF), the reciprocal of PRT.

Recently, frequency-hopping (FH) MIMO (FH-MIMO) radar has attracted extensive interest in DFRC designs \cite{7944485,DFRC_AmbiguityFunc2018Amin,DFRC_DPSK2019Amin,FH_MIMO_Radar2019_RadarConf,DFRC_FHcodeSel2018,9145282,kai_Accurate_Channel_Estimation,2021_kw_FH_MIMO_Mult_ant,Kai_overview_fhMIMO_dfrc2021,Kai_secureFH_MIMO_DFRC}. 
Compared with other pulsed MIMO radars, FH-MIMO radar further divides each pulse into multiple sub-pulses, also called hops, enabling the communication symbol rate to exceed radar PRF \cite{Kai_overview_fhMIMO_dfrc2021}. Moreover, FH-MIMO radars also provide new DoF for information modulation, e.g., the combinations of hopping frequencies \cite{DFRC_FHcodeSel2018} and also the permutations \cite{Kai_secureFH_MIMO_DFRC}. 

However, as in sole wireless communications, the effective demodulation of FH-MIMO DFRC generally requires accurate channel estimation as well as transceiver time/frequency synchronization. Communication training for FH-MIMO DFRC is only studied by a few works in the literature. In \cite{9145282,kai_Accurate_Channel_Estimation}, the estimations of communication channel and sampling timing offset (STO) are studied for FH-MIMO DFRC with a single antenna receiver equipped at the communication user end (UE). In \cite{2021_kw_FH_MIMO_Mult_ant}, the deep fading issue that can severely degrade DFRC performance is identified and solved by introducing multi-antenna receiver for UE. 
Novel waveforms and methods are also designed to estimate channel and timing offset.
Despite the effectiveness of these designs \cite{9145282,kai_Accurate_Channel_Estimation,2021_kw_FH_MIMO_Mult_ant}, they ignored the carrier frequency offset (CFO) and other hardware errors, e.g., inconsistency of transceiver front-ends.

In this work, we develop a practical FH-MIMO DFRC scheme by comprehensively treating all hardware errors, channel estimation, as well as time and frequency synchronization. Using software defined ratio (SDR) platforms and commercial off-the-shelf (COTS) products, we build an FH-MIMO DFRC experiment platform with both radar and communications functions. Moreover, we carry out over-the-air experiments outdoors and indoors, validating the effectiveness of the proposed designs and analysis in real-life scenarios. 
The main contributions and results are summarized as follows. 

\begin{enumerate}[leftmargin=*]
	\item We investigate the impact of practical hardware errors on FH-MIMO DFRC, including STO, CFO and front-end errors (FEE). Here, FEE includes the coupled errors from radio frequency (RF) chains and antennas on both radar transmitter and communication receiver sides. We model these errors and unveil their non-trivial impact on FH-MIMO DFRC. To the best of our knowledge, this is the first time all these hardware errors are jointly considered for FH-MIMO DFRC. 
	
	\item We design new DFRC waveforms by introducing moderate changes to conventional FH-MIMO radar waveforms. 
	We also develop a low-complexity algorithm jointly estimating STO, CFO and FEE at a communication receiver. 
	Moreover, we identify some useful features of the impact of STO, CFO and FEE under the proposed waveforms, and exploit the features to further improve the accuracy of estimating these practical errors.

	\item We build a first FH-MIMO DFRC experiment platform based on Xilinx Zynq SDR \cite{MathWorks} and ADI's FMCOMMS3 RF board \cite{FMCOMMS3}. We also conduct first over-the-air experiments, performing both radar and communications in the same time. 
	Using the proposed FH-MIMO DFRC waveforms, the radar sensing results highly match the sensing scenario, as extracted from a high-resolution satellite map. This manifests that the proposed waveform design has a minimal impact on radar sensing. 
	Moreover, we process the experiment data collected at a communication receiver using the proposed estimation methods. The achieved communications performance is greatly improved over prior art that does not consider all hardware errors as we do. 
\end{enumerate}

We underline that, though our work is focused on FH-MIMO DFRC, the design and analysis has the potential to serve DFRC based on other radars and communications systems. This is because the hardware errors considered in this work, namely STO, CFO and FEE, are common to most, if not all, wireless systems.

We also remark that most FH-MIMO DFRC, as well as other radars-based DFRC, have mainly been performed through theoretical analysis and simulations. Only a few works have illustrated DFRC through prototypes or proof-of-concept platforms. In \cite{9413379}, the communications function of the FH-MIMO DFRC using differential PSK modulations is implemented using the universal software radio peripheral (USRP). 
With a focus on validating the communication feasibility, the work employs single-antenna transmitter and receiver, and makes them synchronized. In contrast, we consider a more practical case in our work with far separated transmitter and receiver which are not physically synchronized. 
In \cite{Spatial_Modulation_JRCS}, a prototype is developed to demonstrate a spatial modulation-based DFRC scheme. 
In \cite{FCR70}, a low-complexity proof-of-concept platform named JCR70 was developed for all-digital joint communications radar at a carrier frequency of $ 73 $ GHz and a bandwidth of $ 2 $ GHz. These works \cite{Spatial_Modulation_JRCS,FCR70} employ specially designed hardwares for specific DFRC schemes. Despite a lack of generality, they are pioneers in respective areas. Moreover, we notice that, since around 2009, there has been a constant interest in using SDR platforms to perform communication waveform-based radar sensing \cite{Implementation_GNU_radio_USRP,Exp_OFDM_SDR_GNU_USRP2,SDR_JCS}.
These works provide great guidance in designing proof-of-concept prototypes based on SDR platforms. 
Nevertheless, they mainly focus on using communication waveforms for sensing, while we deal with a different problem of using radar signals for communications.

The rest of the paper is organized as follows. 
Section \ref{sec: FH-MIMO DFRC} provides the signal model of FH-MIMO DFRC and introduces how information is embedded in the DFRC. Section \ref{sec: A novel FH-MIMO DFRC waveform design} first illustrates the impact of practical hardware errors on the FH-MIMO DFRC and then develops new waveforms and methods to estimate and remove those errors. 
Section \ref{sec: test_plateform} builds an FH-MIMO DFRC experiment platform and shows simulation and experiment results. 
Section \ref{sec: Conclusion} concludes the work.

\section{Signal Model of FH-MIMO DFRC}  \label{sec: FH-MIMO DFRC}
This section briefly describes the principle of FH-MIMO radar-based DFRC, including how the radar works and how data communications is performed by reusing the radar waveform.
\subsection{FH-MIMO Radar} \label{subsec: radar system}
The FH-MIMO radar considered here is a pulse-based orthogonal MIMO radar. It uses separate but cohrent transceiver arrays to achieve the extended array aperture. 
It also employs the fast frequency hopping, namely, each radar pulse is divided into mutiple sub-pulses, i.e., hops, and the frequency changes over hops and antennas.
Let $B$ denote the radar bandwidth. The frequency band is divided into $K$ sub-bands. 
The baseband frequency of the $k$-th sub-band is  
$f_k=\frac{\left( \left\lfloor { - \frac{{K}}{2}} \right\rfloor +k \right) B}{K} ~ (k=0,1,\cdots ,K-1).$
The baseband frequency of the $h$-th hop at antenna $m$ is denoted by $f_{hm}$ which can take $ f_{k}~(\forall k\in [0,K-1]) $. 
Denote the total number of hops in a radar pulse as $ H $. Then the signal transmitted by antenna $ m $ in a radar pulse can be given by
\begin{align}\label{eq: transmitted signal p h m}
	{s}_{m}(t) = e^{-\mathrm j 2\pi f_{hm} t},~0 \le t \le T+hT,~h=0,\cdots,H-1,
\end{align}
where $T$ denotes the time duration of a hop (sub-pulse). To facilitate DFRC, we employ the following constraints \cite{DFRC_AmbiguityFunc2018Amin},
\begin{align}       \label {eq: fphm ne fphm'}
	f_{hm}\ne f_{hm'}~(\forall m\ne m', ~\forall h),~~{BT}/{K}\in \mathbb{I}_{+},
\end{align}
where $\mathbb{I}_{+}$ denotes the set of positive integers.
As a result of the above constraints, the signals transmitted by the $M$ antennas at any hop are orthogonal, i.e., $\int _0^T s_{hm}(t)s^*_{hm^{'}}(t) \mathrm{d}t=0$ given $\forall m\ne m'$. 
An FH-MIMO radar receiving processing 
scheme will be presented in Section \ref{subsec: fh-mimo receiving processing}.

\subsection{FH-MIMO DFRC} \label {subsec: com code}
With reference to the DFRC scheme introduced in \cite{9145282}.
The communication information can be conveyed by two ways.
\textit{First}, the transmitted signal in each hop and antenna can be
multiplied by a PSK symbol, as denoted by $ e^{\mathrm j{\varpi}_{hm}}$, where $ {\varpi}_{hm}\in \Omega_J $ $ (J\ge 1) $  and $ \Omega_J=\left\{  0,\frac{2\pi}{2^J},\cdots,\frac{2\pi(2^J-1)}{2^J} \right\} $ is a PSK constellation with the modulation order $J$.
\textit{Second}, the combination of the hopping frequencies at each hop is also used for conveying information, which is referred to as frequency hopping code selection (FHCS) \cite{DFRC_FHcodeSel2018}.
In particular, given $K$ radar sub-bands and $M$ transmitter antennas, there can be $C^{M}_{K}$ numbers of combinations when selecting $M$ out of $K$ sub-bands.
FHCS uses these combinations to convey information bits whose maximum number is $\left\lfloor \mathrm{log}_2(C^{M}_{K}) \right\rfloor$.

For simplicity, we use a single-antenna communication receiver to illustrate information demodulation in FH-MIMO DFRC.
The communication-received signal at hop $h$ is 
\begin{align}\label {eq: comm-received signal}
	s_h(t) = \sum_{m=0}^{M-1}  \beta_{hm}e^{\mathrm j{\varpi}_{hm}}  e^{-\mathrm j2\pi f_{hm} t}  + v(t),
\end{align}
where $ \beta_{hm} $ is the complex gain between $ m $-th transmitter antenna of the radar and the communication receiver, and $v(t)$ denotes AWGN, and $ M $ denotes the number of transmit antennas.

To demodulate information symbols, we need to estimate $\{ f_{hm} ~ \forall m \}$ and ${\varpi}_{hm} ~ (\forall m)$.
Given the constraint (\ref{eq: fphm ne fphm'}), the former can be estimated by detecting the strongest $M$ peaks in the Fourier transform of $y_h(t)$.
In contrast, ${\varpi}_{hm} ~ (\forall m)$ is more difficult to estimate, as we need to know $\beta_{hm}$ first.
Their estimations are studied in \cite{9145282} under ideal conditions that no timing or frequency offset exists between the radar transmitter and communication receiver.
In \cite{kai_Accurate_Channel_Estimation}, timing offset is considered; however, frequency offset is not yet. Moreover, array calibration error has not been considered for FH-MIMO DFRC so far. 
These non-ideal conditions are investigated in the following.

\section{Practical FH-MIMO DFRC Design} \label {sec: A novel FH-MIMO DFRC waveform design}
In this section, we first investigate the impact of practical transceiver errors on FH-MIMO DFRC. 
Then we design waveforms and propose novel methods to estimate and remove the errors.

\subsection{Impact of Practical Transceiver Errors} \label {subsec: err source}
The clock asynchrony between the radar transmitter and a communication receiver can cause STO and CFO. 
Let $\Delta \omega$ denote the CFO. It changes over time slowly and hence can be treated as a fixed value here. 
Different from CFO, STO accumulates, and its impact varies fast over time. Let $\Delta t_0$ denote the initial STO and $\Delta T_s$ be the sampling time difference between the radar transmitter and the communication.
Then at the $h$-hop of the $i$-th PRT, the accumulated STO can be given by
\begin{align}\label{eq: Delta_t_ih}
	\Delta t_{ih} = \Delta t_0 + (i N_{\mathrm{p}} + hN_{\mathrm{h}}) \Delta T_{\mathrm{s}},
\end{align}
where $N_p$ ($N_{\mathrm{h}}$) is the number of samples in a PRT (hop). Based on (\ref {eq: comm-received signal}), the communication-received signal in the $h$-th hop and $i$-th PRT, with CFO and STO included, can be expressed as
\begin{align}\label {eq: s_{ih}(t)}
	\begin{split}
		s_{ih}(t)=\sum _{m=0}^{M-1} \beta _{ihm}(\omega  _{ihm}) \mathrm{rect} \left(  \frac{t-iT_{\mathrm{p}}-hT}{T}   \right) \times \\
		e^{\mj(\omega _{ihm} + \Delta \omega)(t + \Delta t_{ih})}  e^{\mj \varpi _{ihm} },
	\end{split}
\end{align}
where an additional subscript $(\cdot) _{i}$ is used to indicate the PRT index and $ \mathrm{rect}(\frac{x}{T})$ is the rectangular function that takes one for $x \in [0,T]$  and zero elsewhere.
Here, $T_{\mathrm{p}}$ and $T$ denote the time of a PRT and a hop, respectively. 
Different from $\beta_{hm}$ in (\ref{eq: comm-received signal}), $\beta_{ihm} (\omega_{ihm})$ in (\ref{eq: s_{ih}(t)}) is a function of $\omega_{ihm}$ to account for other frequency dependent gains caused by the radio frequency chains of different antennas.

Calculating the Fourier transform of $s_{ihm}(t)$ at $\omega = \omega_{ihm}$, we obtain

\begin{align}   \label {eq: S_{ih}(domega,T)}
	& S_{ihm} \overset{(a)} {=} \int_{0}^{T} s_{ihm}(\tilde{t}) e^{-\mj \omega_{ihm} (\tilde{t} + iT_{\mathrm{p}} + hT})  \mathrm{d} \tilde{t} \overset{(b)}{\approx} \int_{0}^{T} \beta_{ihm}(\omega_{ihm})  \nonumber \\	
	& \qquad e^{\mj \left( \omega_{ihm} +  \Delta \omega \right) \left( \tilde{t} + iT_{\mathrm{p}} + hT + \Delta t_{ih} \right) }  
	e^{\varpi _{ihm}}  e^{-\mj \omega_{ihm} \left( \tilde{t} + iT_{\mathrm{p}}  + hT \right)} \mathrm{d} \tilde{t}  \nonumber \\	
	&\overset{(c)}{\approx} \mathcal{A}(\Delta \omega) \beta_{ihm}(\omega_{ihm}) 		
	e^{\mj \varpi_{ihm}} e^{\mj \omega_{ihm} \left(\Delta t_0 + (iN_{\mathrm{p}} + hN_{\mathrm{h}})\Delta T_{\mathrm{s}} \right)} \times   \nonumber \\
	& \qquad e^{\mj \Delta \omega (iT_{\mathrm{p}} + hT)}   e^{\mj \Delta \omega  (iN_{\mathrm{p}} +hN_{\mathrm{h}}) \Delta T_{\mathrm{s}}  }
\end{align}
where the substitution $\tilde{t} = t-iT_{\mathrm{p}} - hT$ is performed to get $\overset{(a)}{=}$ with the integral variable changed from $t$ to $\tilde{t}$; 
the expression of $s_{ihm}(t)$ given in (\ref {eq: s_{ih}(t)}) is plugged in $\overset{(a)}{=}$ to get $\overset{(b)}{\approx}$; and $\overset{(c)}{\approx}$ is obtained by replacing $\Delta t_{ih}$ with its expression given in (\ref {eq: Delta_t_ih}) and by taking $e^{\mj \Delta \omega  \Delta t_0} \approx 1$. Note that the integral over $\tilde{t}$ yields
\begin{equation}   \label {eq: mathcal{A}}
	\begin{split}
		\mathcal{A}(\Delta \omega) = T \mathrm{sin} \frac{\Delta \omega T}{2} \Big/ \left( \frac{\Delta \omega T}{2}  \right) e^{\mj \frac{\Delta \omega T}{2}}.
	\end{split}
\end{equation}
Moreover, the approximation $\overset{(b)}{\approx}$ is because we have neglected the Fourier transforms of the signals from other antennas.

It is obvious from (\ref{eq: S_{ih}(domega,T)}) that STO, as indicated by $\Delta t_{ih}$ and CFO, as represented by $\Delta \omega$, have non-trivial impact
on communication demodulation. 
In order to estimate the PSK symbol $\varpi_{ihm}$, all the other phases need to be estimated and suppressed first. This is studied next. In particular, we start with developing the demodulation methods, during which we shall introduce some conditions on the waveform to enable the new methods. Then, we translate those conditions to waveform design.

\subsection{Proposed Demodulation Method} \label{subsec: proposed waveform}

From (\ref {eq: S_{ih}(domega,T)}), we obtain the following, when $\omega_{ihm} =0$ and $\varpi_{ihm}=0$, 
\begin{equation}   \label {eq: S tilde ihm}
	\begin{split}
		\tilde{S}_{ihm} = {S}_{ihm}\mid _{\omega_{ihm} = 0}  = \mathcal{A}(\Delta \omega) \beta_{ihm}(0) \times \\
		e^{\mj \Delta \omega (iT_{\mathrm{p}} + hT) }   e^{\mj \Delta \omega (iN_{\mathrm{p}} + hN_{\mathrm{h}}) \Delta T_{\mathrm{s}}}   .
	\end{split}
\end{equation}
\textit{Note that $ i $, $ h $ and $ m $ are indexes of PRT, hop and antennas, respectively.} 
The result in (\ref{eq: S tilde ihm}) facilitates the estimation of $\Delta \omega$, as detailed below. 

First, similar to $ \tilde{S}_{ihm} $, we can set $ \omega_{(i+1)hm} = 0 $ and obtain
\begin{align}   \label {eq: S tilde i+1 hm}
	& \tilde{S}_{(i+1)hm} = {S}_{(i+1)hm}\mid _{\omega_{(i+1)hm} = 0}  = \mathcal{A}(\Delta \omega) \beta_{(i+1)hm}(0) \times  \nonumber \\
	& ~~~~ e^{\mj \Delta \omega ((i+1)T_{\mathrm{p}} + hT) }   e^{\mj \Delta \omega ((i+1)N_{\mathrm{p}} + hN_{\mathrm{h}}) \Delta T_{\mathrm{s}}}.
\end{align}
Then, taking the ratio between $\tilde{S}_{(i+1)hm}$ and $\tilde{S}_{ihm}$ leads to
\begin{equation}   \label {eq: widetilde{S1}/S}
	\begin{split}
		\frac{\tilde{S}_{(i+1)hm}}{\tilde{S}_{ihm}} = \frac{\beta_{(i+1)hm}(0)}{\beta_{ihm}(0)} e^{\mj \Delta \omega (T_{\mathrm{p}} + N_{\mathrm{p}} \Delta T_{\mathrm{s}})}
		\approx e^{\mj \Delta \omega T_{\mathrm{p}}},
	\end{split}
\end{equation}
where the approximation is because $T_{\mathrm{p}} \gg N_{\mathrm{p}} \Delta T_{\mathrm{s}} $ and $\beta_{ihm}(0) \approx \beta_{(i+1)hm}(0)$.
The validity of the first condition is illustrated in \ref{app: Tp>> N Delta Ts}. For the second, it is because the
channel is approximately unchanged in a single PRT with a short time duration, e.g., 40 $\mu$s to be validated in our experiment.
From (\ref {eq: widetilde{S1}/S}), we can estimate CFO as
\begin{equation}   \label {eq: Delta omega estimate}
	\begin{split}
		\widehat{\Delta \omega} = \mathrm{arg}\left\{ \frac{\tilde{S}_{(i+1)nm}}{\tilde{S}_{ihm}} \right\}
		\Big/ T_{\mathrm{p}} .
	\end{split}
\end{equation}

Note that $\Delta \omega$ and $\omega$ are the same multiples of $\Delta f_{\mathrm{CLK}}$ and $f_{\mathrm{CLK}}$, respectively. Here, $\Delta f_{\mathrm{CLK}}$ denotes the clock offset and $f_{\mathrm{CLK}}$ is the nominal clock frequency. The ratio between $\Delta f_{\mathrm{CLK}}/f_{\mathrm{CLK}}$ is often called the clock stability. Therefore, with $\widehat{\Delta \omega}$ attained, we can estimate the clock stability, as given by
\begin{equation}   \label {eq: rho estimate}
	\begin{split}
		\widehat{\rho} = \widehat{\Delta \omega} \big/ \omega,
	\end{split}
\end{equation}
where $\omega$ denotes the nominal local oscillator angular frequency. 
Based on (\ref {eq: Ts}) in Appendix \ref {app: Tp>> N Delta Ts}, we can further estimate STO $\Delta T_{\mathrm{s}}$ as
\begin{equation}   \label {eq: Delta Ts estimate}
		\widehat{\dts} = - \widehat{\rho}\big/ \left(f_{\ms}^{\mt}(1-\widehat{\rho}) \right),
\end{equation}
where $f_{\ms}^{\mt}$ is the sampling frequency at the transmitter. 
With $\dts$ estimated, we see from (\ref{eq: Delta_t_ih}) that $\dtih$ is also partially
estimated.

From (\ref {eq: S_{ih}(domega,T)}), we see that the remaining unknowns that hinder communication demodulation, i.e., the estimation of $ \varpi_{ihm} $, is $\beta_{ihm}(\oihm)e^{\mj \oihm \dtzero}$. 
The coupling of the two terms makes their individual estimates difficult to obtain. Thus, we consider their joint estimation.
To do so, we introduce $ \breve{S}_{i_1h_1m}~(\forall i_1,~\forall h_1\ne h) $ which is obtained by taking $ \varpi_{i_1h_1m}=0$ and $ \omega_{i_1h_1m}=2\pi k/K $ in (\ref{eq: S_{ih}(domega,T)}). 
{Different from $ \tilde{S}_{ihm} $
given in (\ref{eq: S tilde ihm}) with the zero hopping frequency, $ \breve{S}_{i_1h_1m} $ is obtained under non-zero hopping frequency. Moreover, they can be attained under the same PRT with $ i_1=i $, but they are always obtained in different hops, i.e., $ h_1\ne h $.}

Assuming the above conditions are satisfied, let us check the ratio between $ \breve{S}_{i_1h_1m} $ and $ \tilde{S}_{i_1hm} $, where the latter is obtained by taking $ i=i_1 $ in (\ref{eq: S tilde ihm}). The ratio can be expressed as
\begin{align}  \label {eq: d_mk}
	&\md _{mk} = \frac{\breve{S}_{i_1 h_1 m}}{\tilde{S}_{i_1 h m}}
	=\mathcal{C} \frac{{\beta}_{i_1 h_1 m}(\frac{2\pi k}{K})}    {{\beta}_{i_1 h m}(0)}  e^{\mj \frac{ 2\pi k \Delta t_0}{K}} , ~k=1,\cdots,K-1 \\
	&\mathrm{s.t.} ~\mathcal{C} = e^{\mj \frac{2 \pi k (i_1N_{\mpp} + h_1N_{\mathrm{h}}) \dts}{K}} 
	e^{\mj \Delta \omega(h_1 -h)T}   e^{\mj \Delta \omega(h_1 -h) N_{\mathrm{h}} \dts}. \nonumber
\end{align}
Similarly, let us further construct the ratio between $\breve{S}_{i_2 h_2 m}$ and $\tilde{S}_{i_2 h m}$ with $ i_2\ne i_1 $ and $ h_2\ne h $. 
After some basic calculations, we attain
{\begin{align}   \label {eq: S i_2h_2m/S i_2hm}
	& \frac{\breve{S}_{i_2 h_2 m}}{\tilde{S}_{i_2 h m}}=\md_{k}^{\prime} \mathcal{D} e^{\mj \varpi_{i_2 h_2 m}}, ~\mathrm{s.t.}~ \md_{k}^{\prime} =\mathcal{C} \frac{\beta_{i_2 h_2 m} \left(\frac{2 \pi k}{K} \right)} {\beta_{i_2 h m}(0)}  e^{\mj \frac{2 \pi k \dtzero}{K}}, \nonumber \\
	&  \mathcal{D} = e^{\mj \frac{2 \pi k \left( (i_2-i_1)N_{\mpp} + (h_2-h_1)N_{\mathrm{h}} \right) \dts}{K}} 
	e^{\mj \Delta \omega (h_2 -h_1)T} \times  \nonumber\\
	& \qquad e^{\mj \Delta \omega (h_2-h_1) N_{\mathrm{h}}\dts},
\end{align}}
where $\mathcal{C}$ is given in (\ref{eq: d_mk}), and $ h_1 $ in $ \mathcal{D} $ is due to the inclusion of $ \mathcal{C} $ in $ \md_k' $. 
Note that $\mathcal{D}$ can be estimated based on the estimates obtained in
(\ref {eq: Delta omega estimate}) and (\ref 
{eq: Delta Ts estimate}). 
Assuming that $ \md_k' $ is known for the moment, we can then estimate $\varpi_{i_2 h_2 m}$ as
\begin{equation}   \label {eq: demodulation}
	\widehat{\varpi}_{i_2 h_2 m} = \mathrm{arg} \left\{  \frac{\breve{S}_{i_2 h_2 m}}   {{\md_{k}^{\prime}} \mathcal{D} \tilde{S}_{i_2 hm}}    \right\} 
	~\mathrm{s.t.}~ \omega_{i_2 h_2 m} = 2\pi k /K.
\end{equation}

Our next question is how to know $ \md_k' $. Comparing (\ref{eq: d_mk}) and (\ref{eq: S i_2h_2m/S i_2hm}), we see that $ \md_k' $ has a very similar form to $ \md_{mk} $. In fact, they are approximately the same, as ensured by the following lemma. 

\begin{lemma} \label{lm: dk'=dmk}
	Provided that $| (i_1-i_2)T_{\mpp} |$ is smaller than the stable time of the transceiver front-ends, we have $\md_{k}^{\prime} = \md_{mk}$, where $T_{\mpp}$ is the PRT duration. 
\end{lemma}

\begin{IEEEproof} \label{rmk: dk'=dk}
{From (\ref{eq: d_mk}) and (\ref {eq: S i_2h_2m/S i_2hm}), we see that $\md_{k}^{\prime}$ and $\md_{mk}$ are almost the same other than some differences in the subscripts of the $\beta _{\cdot} (\cdot)$ terms. As illustrated in the texts below (\ref{eq: s_{ih}(t)}), $\beta _{\cdot} (\cdot)$ is the composite impact of the channel response and the complex gains of the transceiver front-ends.
	In the same radar pulse, the channel response can be seen fixed. Thus, the two ratios $\frac{\beta_{i_1 h_1 m} \left(\frac {2\pi k }{K} \right) }{\beta_{i_1 h m}(0)}$ and $\frac{\beta_{i_2 h_2 m} \left(\frac {2\pi k }{K} \right) }{\beta_{i_2 h m}(0)}$  are only dependent on the complex gains of transceiver front-ends.
	As a result, the ratios are the same if $i_1$ and $i_2$ satisfy the condition stated in the lemma. We notice that in modern transceivers, front-ends are generally stable in a contiguous operation, i.e., a whole course of running after a system is powered on. 
	This is also validated through our experiments, as to be presented in Section \ref{subsec: Experiment}.}
\end{IEEEproof}

\begin{algorithm}
	\small
	\captionof{algorithm}{\small Proposed FH-MIMO DFRC Scheme\label{alg: proposed demodulation method}}
	
	\textit{Input}: $ M $ (radar transmitter antenna number), $ H $ (the number of hops in a radar pulse), $ T $ (hop duration), $ T_{\mathrm{p}} $ (PRT duration), $ K $ (the number of sub-bands), $ B $ (radar bandwidth; also the sampling frequency), $ s_{ih}(t) $ given in (\ref{eq: s_{ih}(t)}) (the time-domain communication-received signal)

	\begin{enumerate}[leftmargin=*]
		
		\item For each $ i $ in $ \mathcal{S}_x = \big\{ (0,1,\cdots,K-1)+xK~(\forall x\ge 0)\big\} $:
		
		\begin{enumerate}
			\item Take the Fourier transform of $ s_{ih}(t) $;
			
			\item Identify the largest $ M $ peaks, yielding $ S_{ihm} $ given in (\ref{eq: S_{ih}(domega,T)});
			
			\item For $ \forall m $, calculate $ \md_{mk} $ given in (\ref{eq: d_mk}) by taking $ i_1=i $, $ h_1=m+1 $ (due to (D2)) and $ h=m $ (due to (D1));
			
			\item For $ \forall h_2,m $, Calculate $ \frac{ \breve{S}_{i_2h_2m} }{ \tilde{S}_{i_2hm} } $ in (\ref{eq: S i_2h_2m/S i_2hm}) by taking $ i_2=i $ and $ h=m $ (due to (D1)); 
		\end{enumerate}
		
		\item Estimate $ \widehat{\domega} $ as in (\ref{eq: Delta omega estimate}), where $ h=m $ based on (D1); 
		
		\item Estimate $ \widehat{\dts} $ jointly using (\ref{eq: rho estimate}) and (\ref{eq: Delta Ts estimate}); 
		
		\item For $ \forall i\in \mathcal{S}_x $, $ \forall m $, $ \forall h\ne m $ or $ (m+1) $:
		
		\begin{enumerate}
			\item If $ \oih=2\pi k/K $, set $ i_1=xK+k $, $ h_1=m+1 $, $ i_2=i $ and $ h_2=h $;					
			\item Estimate $ \mathcal{D} $ based on (\ref{eq: S i_2h_2m/S i_2hm});	
			\item Estimate $ \varpiIhm $ based on (\ref{eq: demodulation}) with $ h=m $ taken for $ \tilde{S}_{i_2hm} $ in the denominator;
		\end{enumerate}			
		
		\vspace{-3mm}
		
	\end{enumerate}
\end{algorithm}

\subsection{Novel DFRC Waveform Designs} \label{subsec: waveform design}

We have shown above that under certain conditions imposed on FH-MIMO waveforms, we can suppress unknown channel and hardware errors to estimate communication symbols. 
Those conditions can be ensured through proper waveform designs, as illustrated below.

From (\ref {eq: S tilde ihm}) to (\ref {eq: Delta Ts estimate}), we can see the importance of the zero baseband frequency, i.e., $k=0$ for some $ \omega_{ihm} $. 
Since different antennas have distinct channel responses and front-end gains, we need to ensure that each antenna takes the zero baseband frequency at least once. 
Achieving this will require at least $M$ hops,
since different hopping frequencies are required to used for different antennas in the same hop, as enforced in (\ref{eq: fphm ne fphm'}). 
Thus, our first waveform design can be established as

\vspace{2mm}

\noindent{\it Design 1 (D1): $ \oih=0 $ and $ \varpiIhm=0 $ at $ h=m $, given $ \forall i $.}

\vspace{2mm}

From Lemma \ref{lm: dk'=dmk}, we know that $ \md_{mk} $ needs to be computed for estimating $ \md_k' $. From (\ref{eq: d_mk}), we see that $ \md_{mk} $ is obtained under $ \oih=2\pi k/K $ and $ \varpiIhm=0 $. Thus, to avoid heavily changing the originally radar waveform, we adopt the following waveform design to calculate $\md_{mk}~(k=0,1,\cdots,K-1) $ over $ K $ different PRTs:

\vspace{2mm}

\noindent{\it Design 2 (D2): $ \oih=2\pi\myModulo{i}{K} $/K and $ \varpiIhm=0 $ at $ h=m+1 $, given $ H\ge M+1 $, where $ \myModulo{i}{K} $ denotes the modulo-$ K $ of $ i $.}

\vspace{2mm}

\noindent Note that $ h=m+1 $ is because $ h=m $ has been occupied in Design 1.

The two designs are sufficient for effective communication demodulation in a practical FH-MIMO DFRC with hardware errors and unknown channels. For clarity, we summarize the whole procedure in Algorithm \ref{alg: proposed demodulation method}. While most steps in Algorithm \ref{alg: proposed demodulation method} are straightforward based on the illustrations in this section, we provide some more notes on several key steps. From Step 1), we see that every consecutive $ K $ PRTs are jointly used for communication demodulation. The main reason is that (D2) only allows us to estimate one $ \md_{mk} $ per PRT. While this design is not a must, it introduces minimal changes to the primary radar function. 
In Step 1b), identifying $ M $ peaks from the Fourier transform result is not hard; however, assigning the peaks to the $ M $ radar transmitter antennas. 
Thus, we employ another waveform constraint \cite{kai_Accurate_Channel_Estimation}
\begin{align}
	\omega_{ihm}<\omega_{ihm'}~\forall i,h~\mathrm{and}~\forall m'>m.
\end{align}
To implement the constraint, we let the FH-MIMO radar select its hopping frequencies randomly, then simply re-order the frequencies in ascending order, and assign them to the antennas, one each. A nice feature was disclosed in \cite{9145282}, stating that the above re-ordering does not change the range ambiguity function of the underlying FH-MIMO radar. In Step 2) of Algorithm \ref{alg: proposed demodulation method}, the estimates $ \widehat{\domega} $ obtained under different $ i $'s can be averaged to improve the estimation performance. Then Step 3) can be performed based on the improved $ \widehat{\domega} $ to get a more accurate $ \widehat{\dts} $.

We remark that the FH-MIMO DFRC design is radar-centric in this work; namely, we seek to introduce only minimal changes to the radar yet facilitating effective communications in the presence of hardware errors. The waveform design illustrated above only requires a few hops over antennas to use assigned hopping frequencies, in contrast to random selection in the original radar. Thus, we expect the introduced waveform design has little impact on the radar function. This will be validated in Section \ref{sec: test_plateform}. The significance of our design to communications is that the practical hardware errors are, for the first time, modeled and effectively suppressed for FH-MIMO DFRC. Without considering these inevitable pratical errors, communications performance can be rather poor in practice; this will be demonstrated in Section \ref{sec: test_plateform} through over-the-air experiments.

\subsection{FH-MIMO Radar Receiving Processing} \label{subsec: fh-mimo receiving processing}
Here, we briefly describe an FH-MIMO radar processing scheme. It will be performed in simulations and experiments. 
Let $\mathbf{s}_i(t)=[{s}_{i1}(t),{s}_{i2}(t),\cdots ,{s}_{iM}(t)]^{\rm T}$ collect the signals transmitted by the $M$ antennas in the $ i $-th PRT, where $ {s}_{im}(t)~(\forall m) $ is obtained by replacing $ 2\pi f_{hm} $ in (\ref{eq: transmitted signal p h m}) by $ \omega_{ihm} $ designed in Section \ref{subsec: waveform design}. 
Also, let 
$\mathbf{y}_i(t)=[{y}_{i1}(t),{y}_{i2}(t), \cdots ,{y}_{iN}(t)]^{\rm T}$ denote the signals received by the $ N $ receiver antennas. 
Denote the steering vectors of the transmitter and receiver arrays by $\mathbf{a}_{\rm t}(\theta)$ and $\mathbf{a}_{\rm r}(\theta)$, respectively. Being co-located, they have the same direction.
Considering a single target for simplicity, the radar echo signal in the $ i $-th PRT can be given by 
\begin{align}\label {eq: receive signal vactor}
	& \mathbf{y}_i(t)=\delta \mathbf{a}_{\rm r}(\theta)  \mathbf{a}^{\rm T}_{\rm t}(\theta) \mathbf{s}_i(t-\tau) +\mathbf{w}_i(t),  \nonumber\\
	& \mathrm{s.t.}~t\in (i-1)T_{\mathrm{PRT}} +[HT,T_{\mathrm{PRT}}],
\end{align}
where $\tau$ represents the target echo delay, $\delta$ denotes the target scattering coefficient,
$\mathbf{w}_i(t)$ collects additive white Gaussian noises (AWGNs), and $ T_{\mathrm{PRT}} $ is the time duration of a PRT. 
As mentioned earlier, the FH-MIMO of interest is a pulsed radar. Thus, there shall be no valid echo signal during the radar transmission, as the receiver is either turned off or saturated with invalid echo signals.  
This explains the starting time of each PRT given in (\ref{eq: receive signal vactor}).
The common steps of the radar receiving processing, as will be performed in simulations and experiments, are described below.

\textit{Matched filtering} is a typical first step of pulsed radar signal processing \cite{book_richards2010ModernRadarPrinciples}. 
The filter coefficients are {$s_{im}^*(-t)~(m=0,1, \cdots ,M-1)$, where ``$()^*$'' takes conjugate.
	As each antenna receives a combination of all transmitted signals, $y_{in}(t) ~(\forall n)$ needs to pass each of the $M$ filters, where $ y_{in}(t) $ denotes the $ n $-th entry of $ \mathbf{y}_i(t) $ given in (\ref{eq: receive signal vactor}). The matched filtering result can be written as,
	\begin{align} \label {eq: y tilde ip (t) conv results}
		\tilde{y}_{ip}(t) = y_{in}(t) \circledast s_m^*(-t) ,~p=(n-1)M+m, 
	\end{align}
	where $ \circledast $ calculates the linear convolution. 
	
	\textit{Moving target detection (MTD)} is often performed after the matched filter. 
	For $ \tilde{y}_{ip}(t)~(\forall p,t) $, a Fourier transform can be performed over $ i $, leading to the so-called range-Doppler map (RDM),
	\begin{align} \label{eq: MTD}
		\tilde{Y}_{fp}(t) = \sum_{i=0}^{I-1} \tilde{y}_{ip}(t) e^{-\mj 2\pi f iT_{\mathrm{PRT}}},
	\end{align} 
	where $ f $ and $t$ span the Doppler and range dimensions, respectively.

	\textit{Target detection} can be performed based on $ \sum_{p=0}^{P-1} | \tilde{Y}_{fp}(t)| $, where the incoherent accumulation over $p$ (indexing spatial channels) is performed as we do not have angle information yet; otherwise, the coherent beamforming can be performed. The constant false-alarm rate (CFAR) detector has been popularly employed for target detection and hence will be used for our simulation and experiment. Interested readers are referred to \cite{book_richards2010ModernRadarPrinciples} for the details of the CFAR detector. An intuitive simulation tutorial is also available at \cite{matlab_cfar}.
	Let $ (f^*,t^*) $ denote the location of a target. Extracting the signal at each RDM, we obtain
	\begin{align}
		\mathbf{z} = \left[  \tilde{Y}_{f^* 0}(t^*) , \tilde{Y}_{f^* 1}(t^*),\cdots,\tilde{Y}_{f^* (P-1)}(t^*) \right]^{\mathrm{T}},
	\end{align}%
	where $ P=MN $ according to (\ref{eq: y tilde ip (t) conv results}).

	\textit{Angle estimation} is carried out using $ \mathbf{z} $. Based on (\ref{eq: receive signal vactor}), the steering vector depicting the spatial information in $ \mathbf{z} $ can be written as $ \tilde{\mathbf{a}}(\theta) = \mathbf{a}_{\rm r}(\theta) \otimes  \mathbf{a}_{\rm t}(\theta) $, where $ \otimes $ denote the Kronecker product. 
	In practice, the radio frequency chains associated with transmitter and receiver antennas always present some differences, as depicted by $\mathbf{e}_{\mathrm{t}}$ and $\mathbf{e}_{\mathrm{r}}$, respectively.
	Incorporating them, the steering vector can be revised as  $ \breve{\mathbf{a}}(\theta)=\tilde{\mathbf{a}}(\theta)\odot (\mathbf{e}_{\mathrm{r}}\otimes \mathbf{e}_{\mathrm{t}}) $. Therefore, a naive angle estimate can be obtained as
	\begin{align}  \label{eq: es_theta}
		\hat{\theta} = \mathrm{argmax}_{\theta_l=2\pi l/L}~~ |\breve{\mathbf{a}}(\theta_l)^{\mathrm{H}} \mathbf{z}|^2,~ l=0,1,\cdots,L-1
	\end{align}
	where $ L $ can take a relatively large value for a fine spatial resolution. More advancing methods, such as the multiple signal classification (MUSIC) \cite{MUSIC} and the DFT interpolation-based methods 	\cite{Kai_padeFreqEst2021TVT}, can be employed for more accurate angle estimation accuracy.

\section{OVER-THE-AIR FH-MIMO DFRC EXPERIMENTS} \label{sec: test_plateform}
\begin{figure}[!t]
	\centering
	\includegraphics[width=1\linewidth]{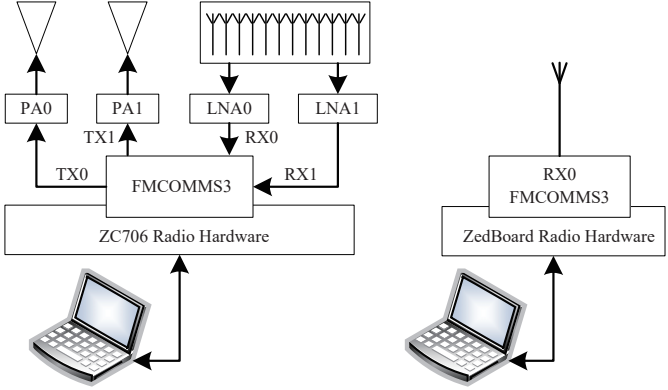}
	\caption{The schematic diagram of the testing system} 
	\label{fig:system_sch}
\end{figure}
In this section, we perform over-the-air experiments to validate the proposed FH-MIMO DFRC scheme.

\subsection{Schematic Diagram and Experiment Platform}  \label{subsec: Schematic Diagram and Experiment Platform}
The schematic diagram of our testing system is shown in Fig. \ref{fig:system_sch}.
We employ the Xilinx Zynq software-defined radio (SDR) ZC706 \cite{ZC706} and ZedBoard \cite{ZedBoard} to build the FH-MIMO radar and communication receiver, respectively. 
Both SDRs are equipped with RF FPGA mezzanine card (RF FMC) boards, FMCOMMS3 \cite{FMCOMMS3} in specific. 
Each FMCOMMS3 supports two transmitting and two receiving RF chains with the RF range of $70$ MHz $\sim$ $6$ GHz and a baseband frequency range of $200$ KHz $\sim$ $56$ MHz. 
A $40$ MHz oscillator with the stability of $10$ ppm is used by each FMCOMMS3 \cite{EpsonTsx3225}. 
Note that both SDRs are supported by MATLAB \cite{MathWorks}. 
Thus, they are connected to host computers, where MATLABs are installed and used to program and control the SDRs independently. 

For the FH-MIMO radar, we use MATLAB to generate the base-band signals of the two transmitting antennas for a CPI and download them to ZC706 once through Ethernet.
The SDR is configured to cyclically transmit the signals. 
The two radar receiving	channels are configured to capture echo signal in a consecutive time of $204.8$ ms, corresponding to $8,192,000$ samples. Also note that the maximum number of samples that can be transferred in one capture is $ 8,388,608 $, which is a limitation of the employed SDR \cite{MathWorks}. 
Note that the echo capture needs to be triggered in MATLAB of the host computer. 
The communication receiver, as configured similar to the radar receiver, also needs a triggering signal from the MATLAB connected to the SDR. 
Next, we provide more elaborations on radar and communication sub-systems.
\begin{figure}[!t]
	\centering
	\includegraphics[width=1\linewidth]{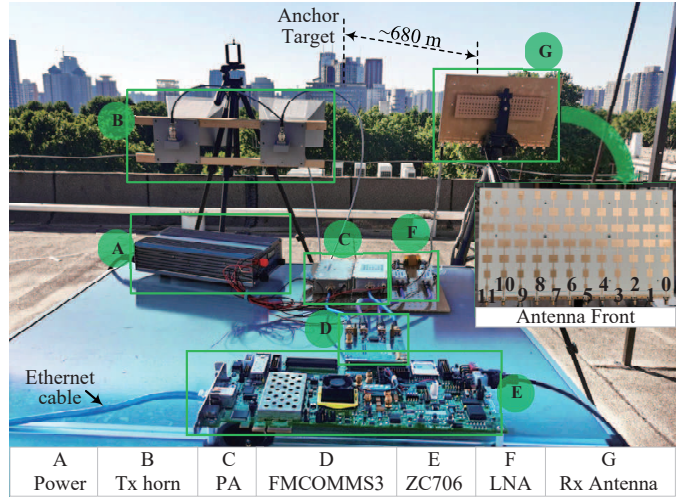}
	\caption{The established FH-MIMO radar, where the host computer is not shown but connected with the SDR through an Ethernet cable.} 
	\label{fig:radar_system}
\end{figure}

\subsubsection{Radar subsystem}
Based on ZC706, we build an FH-MIMO radar platform, as shown in Fig. \ref{fig:radar_system}. 
A host computer (not shown in the figure) is connected to the SDR through an Ethernet cable (which is shown on the lower left side). 
The RF board, FMCOMMS3, is underlain by AD9361 which has the maximum output power of $6.5$ dBm at $5.5$ GHz \cite{AD9361}, where $5.5$ GHz is the carrier frequency used in our experiments. 
Moreover, we use external power amplifiers (PAs) to increase the radar transmission power. 
The maximum output power of the employed PA is $20$ W, i,e., $43$ dBm. By controlling the output power of AD9361, the transmission power is fixed at $2$ W in the sequential experiments.
Moreover, two identical horn antennas are also used for radar transmission, each connected to a RF chain.
For radar receiving, a microstrip uniform linear array of $12$ antennas is used. 
The antenna spacing is half-the-wavelength at the center frequency of $5.5$ GHz. 
Two low-noise amplifiers (LNAs) are used, one for each receiving RF chain on FMCOMMS3. 
Key Parameters of the above components are listed in Table \ref{tab: Parameters of the Established FH-MIMO Radar}.

\begin{table}[!t]  %
	\begin{center}
		\caption{Parameters of the Established FH-MIMO Radar}
		\label{tab: Parameters of the Established FH-MIMO Radar}
		\resizebox{0.5\textwidth}{!}{
		\begin{tabular}{c|c|l}
			\hline
			\textbf{Variable} 	&\textbf{Parameter} 								& \textbf{Value}\\
			\hline
			-					& Central frequency									& $5.5$GHz	\\
			-					& power amplifier gain								& $43$dB	\\
			M					& Number of horn antennas							& $2$	\\
			-					& horn antenna gain									& $20.39$dB	\\
			-					& horizontal beamwidth of horn antenna				& $16^\circ$	\\
			-					& vertical beamwidth of horn antenna				& $15.5^\circ$	\\		
			-					& Maximum transmit power of PA						& $20$W		\\
			N					& Number of radar receiver antennas					& $12$	\\
			-					& Receiving antenna element gain					& $13$dB		\\
			-					& Horizontal beamwidth of microstrip antenna		& $120^\circ$			\\
			-					& Vertical beamwidth of microstrip antenna			& $20^\circ$		\\
			-					& Low noise amplifier gain							& $23$dB		\\
			\hline
		\end{tabular}
	}
	\end{center}
\end{table}

\begin{table}[!t]  %
	\begin{center}
		\caption{Parameters of the FH-MIMO Radar}
		\label{tab: Parameters of the FH-MIMO Radar}
		\resizebox{0.5\textwidth}{!}{
		\begin{tabular}{c|c|l}
			\hline
			\textbf{Variable} 	&\textbf{Parameter} 								& \textbf{Value}\\
			\hline
			$B$					& the Signal Bandwidth								& $20$MHz	\\
			$f_k$				& radar sub-band baseband frequency					& $-10:1:9$ MHz	\\
			$T$					& hop duration    									& $1\mu$s	\\
			$H$					& number of hops per pulse							& $5$	\\
			$T_{\mathrm{p}}$	& PRT												& $40\mu$s  	\\
			$N_{\mathrm{c}}$	& number of PRTs per CPI							& $128$	\\		
			$f_{\mathrm{s}}$	& sampling frequency								& $40$ MHz		\\
			$N_\mathrm{p}$		& number of samples per PRT							& $1600 (= f_{\mathrm{s}}T_{\mathrm{p}}$)	\\
			\hline
		\end{tabular}
		}
	\end{center}
\end{table}

Several notes are given below. \textit{First}, AD9361 has its gain adjustable in $-89.75 \sim 0$ dB for the transmitting RF chain and $0 \sim 61$ dB for receiving RF chain. 
Thus, together with the gains from other components, see Table \ref{tab: Parameters of the Established FH-MIMO Radar}, the maximum transmitting and receiving power gain of the FH-MIMO radar built in Fig. \ref{fig:radar_system} can be {$43 (=0+43)$ dB and $84 (=23+61)$ dB}, respectively. \textit{Second}, during experiments, we place the two horn antennas $6 \lambda$ apart from each other.
According to the MIMO radar processing illustrated in Section \ref{subsec: fh-mimo receiving processing}, the way how the transceiver arrays are placed leads to a virtual array of $2 \times 12 = 24$ antennas. 
\textit{Third}, limited by the number of receiving RF chains on FMCOMMS3, the time division multiplexing (TDM) MIMO is employed to achieve the above-mentioned virtual array. 
As shown in Fig. \ref{fig:radar_system}, the receiving antenna array has $ 12 $ elements, each connected to an SMA port. However, we can see from Fig. \ref{fig:communication_system}, there are only two receiving RF chains. Thus, we collect echo signals from the $ 12 $ antennas in six consecutive data captures. In the $ n $-th $ (n=0,1,\cdots,5) $ capture, the two receiving SMAs are connected to the $ n $-th and $ (n+6) $-th antenna element shown in Fig. \ref{fig:radar_system}. 
Interested readers are referred to \cite{MIMO_radar} for more details on TDM-MIMO radars.

Based on the hardware features illustrated in Section \ref{subsec: Schematic Diagram and Experiment Platform}, we set the parameters of the FH-MIMO radar, as given in Table \ref{tab: Parameters of the FH-MIMO Radar}. 
As the considered FH-MIMO radar is a pulsed radar, the receiver channel suffers from strong self-interference when the transmitter works, leading to a blind zone of $\mathsf{C}HT/2$ (in meter), where $\mathsf{C}$ denotes the microwave speed, $T$ is a hop duration and $H$ is the hop number. Given the limited link budget, see Table \ref{tab: Parameters of the Established FH-MIMO Radar}, the maximum measurable distance of the radar platform would be very limited. 
Thus, we want to keep the blind zone small as well. To do so, we set $T = 1\mu$s and $H = 5$. This leads to a blind zone of $750$ m. 
Other parameters in Table \ref{tab: Parameters of the FH-MIMO Radar} are straightforward based on the descriptions therein.

\subsubsection{Communications Subsystem}
\begin{figure}[!t]
	\centering
	\includegraphics[width=1\linewidth]{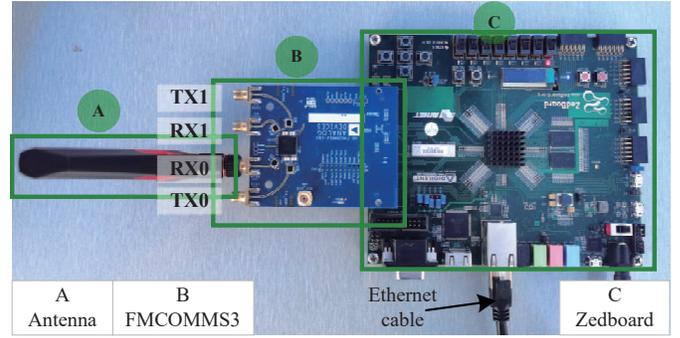}
	\caption{the Communication Subsystem} 
	\label{fig:communication_system}
\end{figure}
It is built on the SDR Zed-Board, FMCOMMS3 and an 8 dBi omnidirectional antenna, as shown in Fig. \ref{fig:communication_system}. 
As only downlink communication is considered in this work, the communication subsystem only receives, hence much simpler compared with the radar system.
We do not use external LNA for the communication subsystem. Thus, its receiving power gain ranges in $0 \sim 61$ dB, solely dependent on AD9361 of FMCOMMS3.

Based on the parameters given in Table \ref{tab: Parameters of the FH-MIMO Radar}, we can calculate the data rate of the radar-enabled communications.
As illustrated in Section \ref{subsec: com code}, the combinations of hopping frequencies are used as communication data symbols. 
Given $K = 20$ sub-bands and $M = 2$ transmitting antennas, we have $\mathrm{C}^2_{20} = 190$. 
Considering the integer number of bits, out of $190$ combinations, $128$ numbers of combinations can be used to convey $7$ bits per radar hop. 
Given $H = 5$, using the combinations convey a total of $7 \times 5 = 35$ bits per PRT. Moreover, PSK is also employed for information demodulation, one symbol per hop and antenna. 
Thus, Considering an $x$-bit PSK modulation. The overall number of bits conveyed by PSK is $xMH = 10x$ per PRF. In summary, the communication data rate is
\begin{align}       \label {eq: communication data rate}
	(35+10x)/T_{\mathrm{p}} = (0.875+0.25x) ~\text{Mbps}.
\end{align}
For $x = 1, 2, 3$ and $4$, the data rate is $1.125, 1.375, 1.625$ and $1.875$ Mbps, respectively.

\subsection{Simulation Analysis} \label{subsec: simulation analysis}
\begin{figure}[!t]     
	\centering
	\includegraphics[width=1\linewidth]{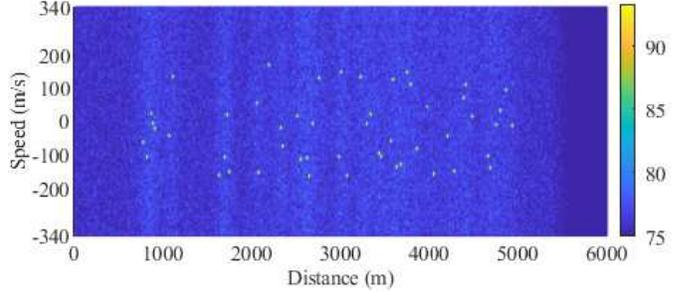}
	\caption{The 50 random targets for simulation.}
	\label{fig: sim_target 50}
\end{figure}
\begin{figure}[!t]     
	\centering
	\includegraphics[width=1\linewidth]{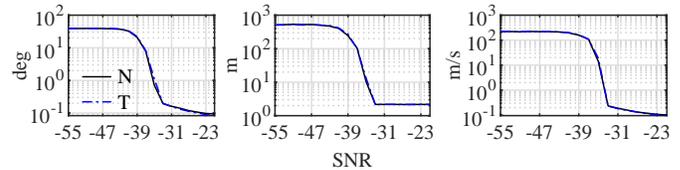}
	\caption{Root mean squared error (RMSE) of angle, distance and velocity	estimations under different SNR, where ``N'' and ``T'' represent new and traditional FH-MIMO radar waveforms, respectively.}
	\label{fig: sim_target Error vs SNR}
\end{figure}
\begin{figure}[!t]     
	\centering
	\includegraphics[width=1\linewidth]{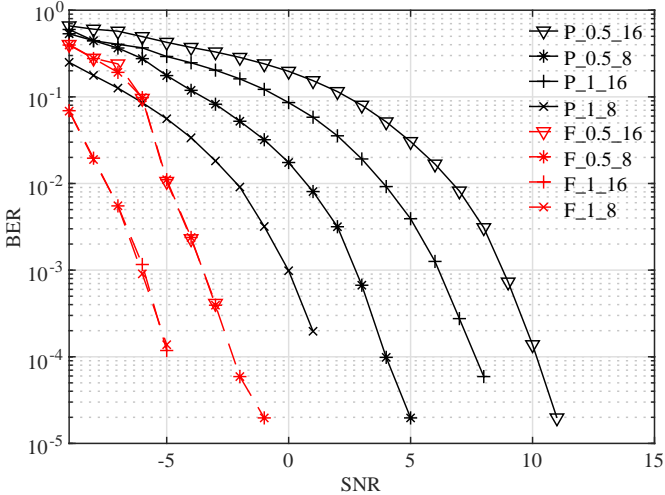}
	\caption{Simulation results of BER curve versus SNR } 
	\label{fig: sim_SNR_BER}
\end{figure}
\begin{figure*}[!t]     
	\centering
	\includegraphics[width=1\linewidth]{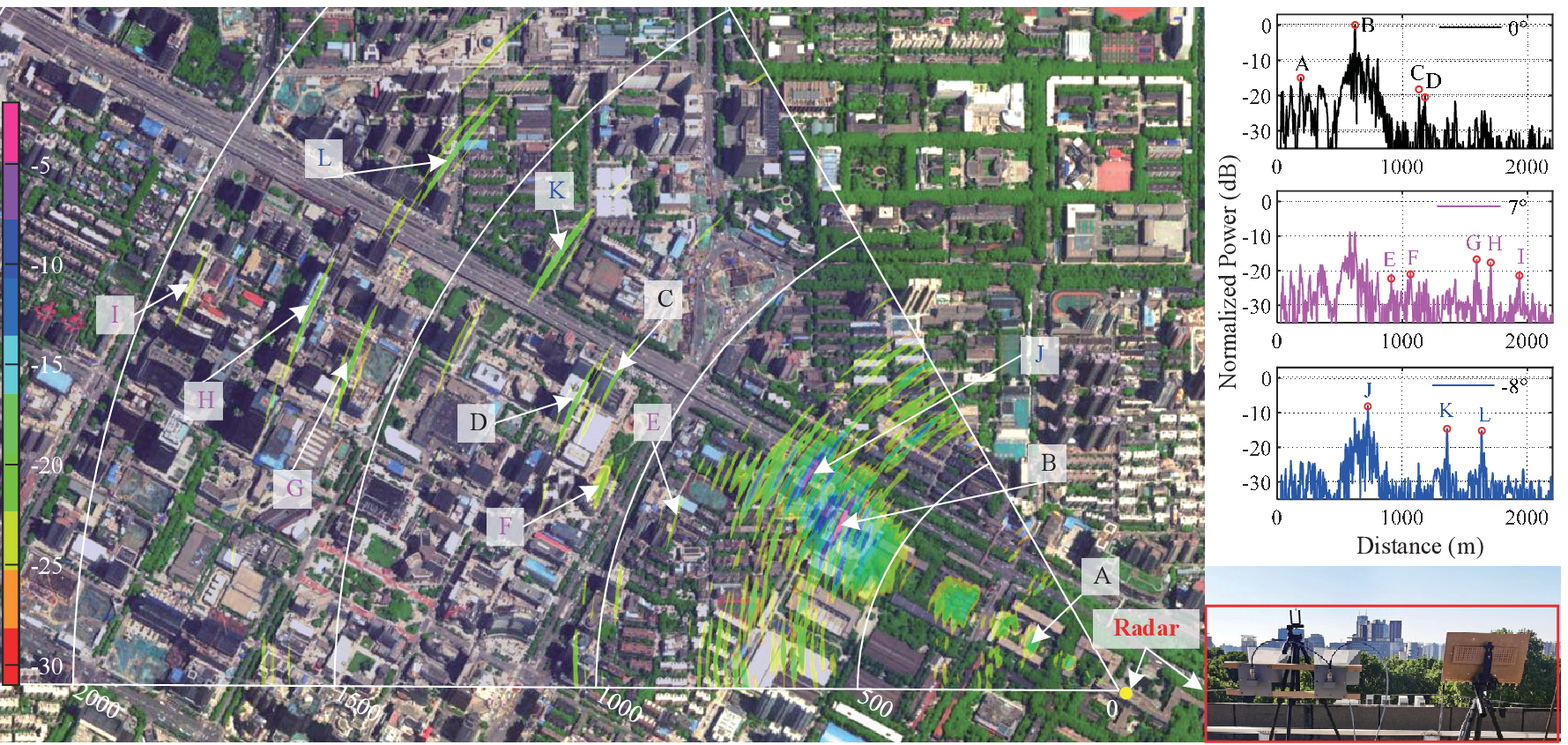}
	\caption{A two-dimensional space-distance radar imaging using the proposed FH-MIMO DFRC waveform and the hardware platform built in Section \ref{subsec: Schematic Diagram and Experiment Platform}.
		The zero-Doppler channel is observed, and hence only static targets are shown in the radar imaging. Representative targets are pinpointed based on the latest
		satellite map of the location \cite{Map}. The distances profiles of three angles covering most pinpointed targets are separately plotted in
		the upper-right sub-figure. The lower-right sub-figure is a trimmed version of Fig. \ref{fig:radar_system}, giving a glimpse on the radar platform and the surrounding environment.} 
	\label{fig: radar target and map1}
\end{figure*}

Before performing experiments, simulations are carried out to validate the proposed FH-MIMO DFRC. 
We start with validating the radar performance, where the radar is configured as per Table \ref{tab: Parameters of the FH-MIMO Radar}.
As for sensing scenario, we set $50$ targets with random speeds, distances  and angles which are uniformly distributed in $[-170, ~170]$ m/s, $[750, ~4,185]$ m and $[-4, ~4]$ deg, respectively. 
To evaluate the root mean squared error (RMSE) of parameter estimations, we perform 100 independent trials, with target parameters randomly generated over trials. 
For each trial, we perform the radar processing illustrated in Section \ref{subsec: fh-mimo receiving processing} for target detection and estimation. Both the conventional FH-MIMO radar in Section \ref{subsec: radar system} and the one modified for DFRC in Section \ref{subsec: waveform design} are simulated for a comparison.

Fig. \ref{fig: sim_target 50} plots a snapshot of a range-Doppler map (RDM) obtained in one trial. We can see many strong points scattering over the range-Doppler domain, each point representing a target. 
As illustrated in Section \ref{subsec: fh-mimo receiving processing}, CFAR is performed based on an RDM. Then the delay and Doppler bins of the detected targets are used to estimate their parameters. 
Fig. \ref{fig: sim_target Error vs SNR} plots the RMSEs of angle, distance and velocity averaged over all targets and trials. 
We see that the RMSEs of all parameter estimates first decrease and then converge, as SNR increases. 
This complies with general understanding, and the convergence is due to the quantized distance, velocity and angle grids used during the estimation; see Section \ref{subsec: fh-mimo receiving processing}. 
More importantly, we see from Fig. \ref{fig: sim_target Error vs SNR} that the traditional and new FH-MIMO radar waveforms lead to almost the same estimation performance.
This validates that the proposed waveform designs for DFRC only incur minimal changes to the underlying FH-MIMO radar.

Next, we demonstrate the communication performance of FH-MIMO DFRC employing the metric of bit error rate (BER). 
FHCS and PSK, as illustrated in Section \ref{subsec: com code}, are simulated. Most radar configurations in Table III are used. 
But here, we also consider two different hop duration, i.e., 0.5$\mu$s and 1$\mu$s. 
As for the PSK, we simulate 8PSK and 16PSK. In this simulation, we let the communication receiver know the channel responses.
Thus, the results here provide a performance lower bound of the experiment results to be presented shortly.

Fig. \ref{fig: sim_SNR_BER} plots the BER performance of different modulations under different settings. We see that FHCS generally has lower BER than PSK modulations. 
This is consistent with previous works, e.g., \cite{9145282}. 
We also see that when the hop duration doubles, both FHCS and PSK achieve better BER performance. This is because the demodulation SNR increases with the hop duration. 
For FHCS, we see from Fig. \ref{fig: sim_SNR_BER} that the modulation order does not affect its BER performance.
This is due to the way FHCS is demodulated. In particular, as illustrated in Section \ref{subsec: com code}, we only need to identify DFT peaks for demodulating FHCS, where the phases of peaks are
irrelevant.

\subsection{Experimental Results}   \label{subsec: Experiment}

\begin{figure}[!t]
	\centering 	
	\subfigure[]{
		\begin{minipage}[t]{0.5\linewidth}
			\centering
			\includegraphics[width=1\linewidth]{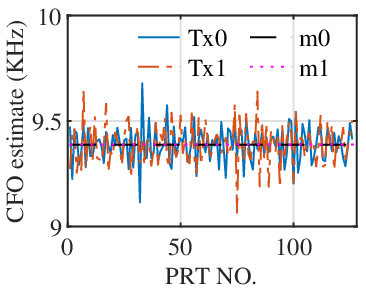}
		\end{minipage}%
	}%
	\subfigure[]{
		\begin{minipage}[t]{0.5\linewidth}
			\centering
			\includegraphics[width=1\linewidth]{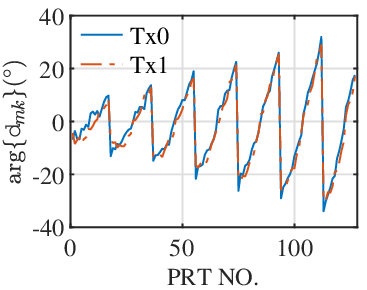}
		\end{minipage}%
	}%
	\centering
	\caption{The CFO  estimate, i.e., $ \widehat{\domega} $, is illustrated in (a); the estimate of $\arg\{\mathsf{d}_{mk}\}$ in (b), where $ \widehat{\domega} $ is obtained in (\ref{eq: Delta omega estimate}) and $\mathsf{d}_{mk}$ in (\ref{eq: d_mk}).
		In a CPI, the value of $k$ cyclic increases with the PRTs, and it can be known from (\ref{eq: d_mk}) that for a $k$, $\arg\{\mathsf{d}_{mk}\}$ increases with the PRT index. 
	}
	\label{fig:communication_Df_DTs}
\end{figure}

\begin{figure}[!t]
	\centering 
	\subfigure[]{
		\begin{minipage}[t]{0.31 \linewidth}
			\includegraphics[width=1\linewidth]{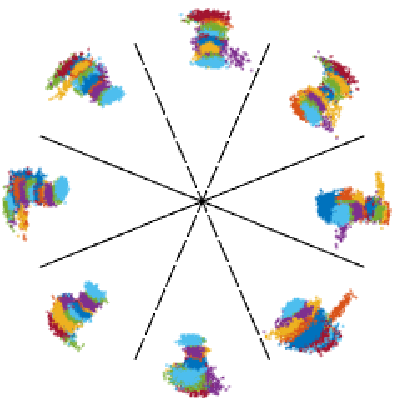}
		\end{minipage}%
	}%
	\subfigure[]{
		\begin{minipage}[t]{0.31\linewidth}
			\includegraphics[width=1\linewidth]{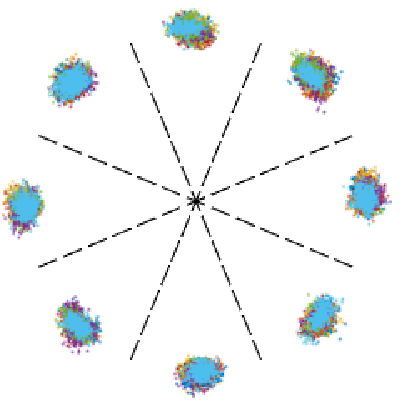}
		\end{minipage}%
	}%
	\subfigure[]{
		\begin{minipage}[t]{0.31\linewidth}
			\includegraphics[width=1\linewidth]{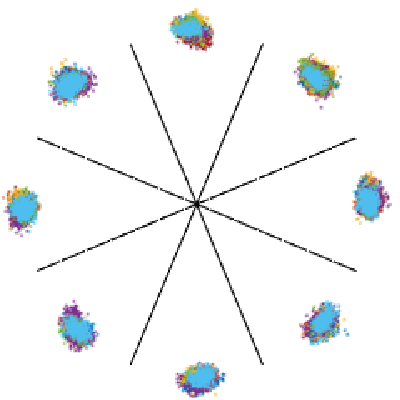}
		\end{minipage}%
	}%

	\subfigure[]{
		\begin{minipage}[t]{0.31\linewidth}
			\centering
			\includegraphics[width=1\linewidth]{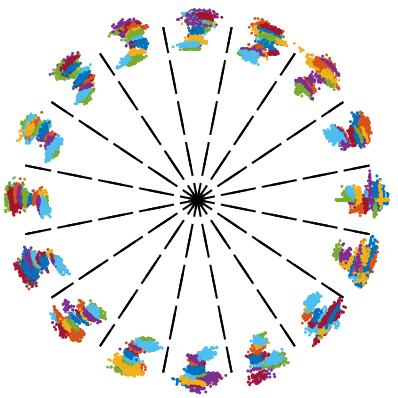}
		\end{minipage}%
	}%
	\subfigure[]{
		\begin{minipage}[t]{0.31\linewidth}
			\centering
			\includegraphics[width=1\linewidth]{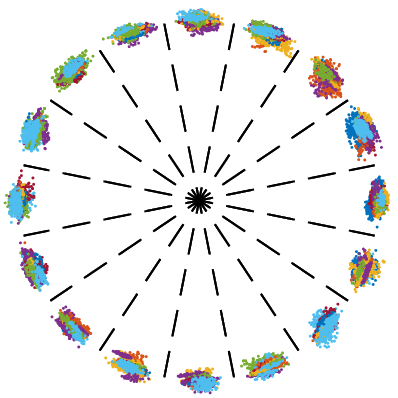}
		\end{minipage}%
	}%
	\subfigure[]{
		\begin{minipage}[t]{0.31\linewidth}
			\centering
			\includegraphics[width=1\linewidth]{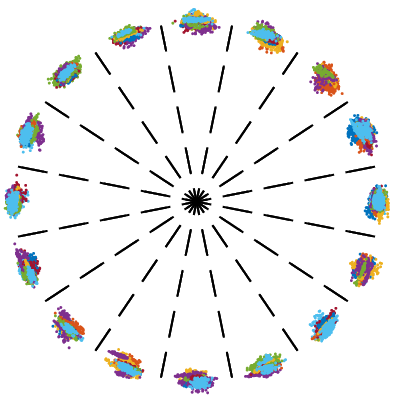}
		\end{minipage}%
	}%
	
	\centering
	\caption{Illustrating demodulation performance, where the top row is for 8PSK and the second row is for 16PSK. The first column is the demodulation results without considering frequency-dependent transceiver gains, which
		approximates the results obtained in \cite{Kai_padeFreqEst2021TVT}. The middle column is based on the proposed design given in Section \ref{sec: A novel FH-MIMO DFRC waveform design}. The third column further averages the estimates of $\mathcal{D}$ over a CPI at the same frequency. }
	\label{fig:communication_decode}
\end{figure}

Employing the hardware platforms illustrated in Section \ref{subsec: Schematic Diagram and Experiment Platform}, we perform over-the-air experiments. 
As shown in Fig.\ref{fig:radar_system}, the radar transceiver is placed on top of a building with the height of about $20$ m. Fig. \ref{fig: radar target and map1} plots the radar imaging results, where the observation distance is up to $3$ km from the radar and the angular region is $[-30^\circ, 30^\circ]$ around the normal direction of the radar. To plot the 2D radar imaging, the zero-th Doppler channel of the MTD result, as obtained in (\ref{eq: MTD}), is extracted in each CPI. 
Then, the beamforming, as illustrated in (\ref{eq: receive signal vactor}), is performed to scan the angular region with a step of $1^\circ$.

To calibrate the radar transceiver arrays, we use a known target (i.e., target B in Fig. \ref{fig: radar target and map1}) to calculate the array calibration coefficients, i.e., $\mathbf{e}_{\mathrm{r}} \otimes  \mathbf{e}_{\mathrm{t}}$ given in (\ref{eq: es_theta}). 
In particular, we place the radar transceiver in such a way that target B is in the normal direction of the radar. 
Since the target distance is known, we extract the signal of the $p$-th $(p = 0, 1, \cdots , P-1)$ virtual spatial channel, i.e., the signal given in (\ref{eq: MTD}), at the known distance and zero-th Doppler bin. 
Given the radar configuration in Table \ref{tab: Parameters of the FH-MIMO Radar}, we have $P(= 24)$ virtual channels. 
Ideally, the extracted signals should be the same. But, as affected by the array calibration errors, their values can be distinct. 
Thus, we use the signal of the first virtual spatial channel as a reference, and all other extracted signals are normalized to the reference one, leading to the array calibration vector. 
In fact, in our experiment, we do not have the facilities to calibrate the radar transceiver arrays. 
Using the above anchor-based method, we are able to obtain a relatively good calibrated array, as demonstrated by  the high match between the measured and map-illustrated targets in Fig. \ref{fig: radar target and map1}.

In the figure, we superimpose the radar imaging above a satellite map of the observed area, where the map is {obtained from \cite{Map}}. 
We see from Fig. \ref{fig: radar target and map1} that the strong signals in the radar imaging match well with the objects observed on the map. 
The distance profiles at the three selected angels further highlight most pinpointed targets. This illustrates the effectiveness of the above-mentioned array calibration. 
More importantly, this validates that the proposed waveform modifications for FH-MIMO radar to accommodate data communications do not obviously affect the primary radar function.

Next, we illustrate the communications performance. In the first set of experiments (i.e., Figs. \ref{fig: radar target and map1}, \ref{fig:communication_Df_DTs} and \ref{fig:communication_decode}), we place the communication receiver, as illustrated in Fig. \ref{fig:communication_system}, about ten meters behind the radar transmitter antennas (which are placed outdoor as shown in Fig. \ref{fig:radar_system}). That is, the communication receiver is in the line-of-sight (LoS) posterior views of radar transmitter antennas. The proposed communication demodulation method, as summarized in Algorithm \ref{alg: proposed demodulation method}, is performed on collected experiment data.

Fig. \ref{fig:communication_Df_DTs}(a) plots the CFO estimate obtained each pair of two consecutive PRTs; see (\ref{eq: Delta omega estimate}), {since there are 128 PRTs, there are 127 results.} We see that there is a non-negligible CFO between the radar transmitter and the communication receiver. Moreover, we also see that the two communication receivers have different CFOs. In addition, we see that the CFO estimate changes over time but stays around approximately a fixed value {(m0 and m1 in Fig. \ref{fig:communication_Df_DTs}(a))}. 
This validates the slow-varying feature of CFO in a certain coherent processing period. This also confirms that we can average the estimates of CFO over a suitable time  period to obtain a more accurate estimation.
{Fig. \ref{fig:communication_Df_DTs}(b) plots the estimate of $\mathsf{d}_{mk}$. We see that $\mathsf{d}_{mk}$ changes over PRTs in the way depicted by (\ref{eq: d_mk}). 
Recall that, in the proposed waveform design, the $i$-th PRT estimates $\md_{mk}$ for $k=(i)_{K-1}$, where $()_{K-1}$ denotes modulo-$(K-1)$.}

Fig. \ref{fig:communication_decode} provides the scatter plots of the demodulated communication symbols, where 8PSK is given in the first row and 16PSK is in the second. 
Three demodulation methods are performed based on the same collected experiment data.
The first method, leading to the results in the first column, neglects the frequency-dependent transceiver gains, which is essentially the method in \cite{kai_Accurate_Channel_Estimation}. 
The second method, as summarized in Algorithm \ref{alg: proposed demodulation method}, leads to the results in the middle column. 
Following the overall steps of Algorithm \ref{alg: proposed demodulation method}, the third method, further averages $\mathcal{D}$ obtained in Step 4b) of Algorithm \ref{alg: proposed demodulation method} in a CPI, 
thus improving the estimate of $\widehat{\varpi}_{i_2 h_2 m}$ and then demodulation performance.
From Fig. \ref{fig:communication_decode}, we see that neglecting frequency-dependent transceiver gains can substantially degrade the communication performance of FH-MIMO DFRC. 
This validates the critical importance of the proposed designs which specifically account for the frequency-dependent transceiver gains. 
From the middle column of Fig. \ref{fig:communication_decode}, we see that the proposed waveform and demodulation method can achieve relatively good communication performance.
{By further averaging $ \arg\{\mathcal{D} \}$ }over time, where $ \mathcal{D} $ is given in (\ref{eq: S i_2h_2m/S i_2hm}), we can further improve the estimation accuracy of $\varpi_{i_2 h_2 m}$  and hence the communication performance, as manifested in the third column of Fig. \ref{fig:communication_decode}.

In the second set of experiment for validating communication performances, we observe the BER performance under different SNRs, as simulated in Fig. \ref{fig: sim_SNR_BER}. 
To precisely control demodulation SNRs, we use two SDRs indoors, one transmitting and one receiving, as illustrated in Fig. \ref{fig: BER_indoor}. The transmitting SDR is the one used for radar transmitter, but is equipped with two omni-directional antennas (with $12$ dBi). During the experiment, we set the gains of the two transmitting RF chains as $0$ dB and adjust the gains of the receiving RF chains to achieve different demodulation SNRs. Fig. \ref{fig: Gain} plots the actual gain under different gain control. The actual gain is estimated by averaging the power of the received signal which is normalized to the estimated gain under the $0$ dB gain control. 

\begin{figure}[!t]    
	\centering
	\includegraphics[width=1\linewidth]{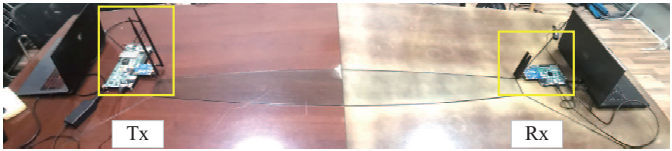}
	\caption{Indoor BER Test Scenario} 
	\label{fig: BER_indoor}
\end{figure}

\begin{figure}[!t]    
	\centering
	\includegraphics[width=1\linewidth]{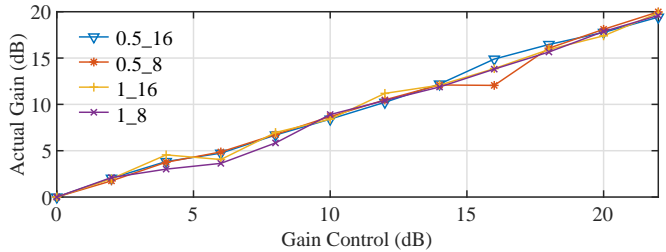}
	\caption{actual gain versus receive gain} 
	\label{fig: Gain}
\end{figure}

\begin{figure}[!t]    
	\centering
	\includegraphics[width=1\linewidth]{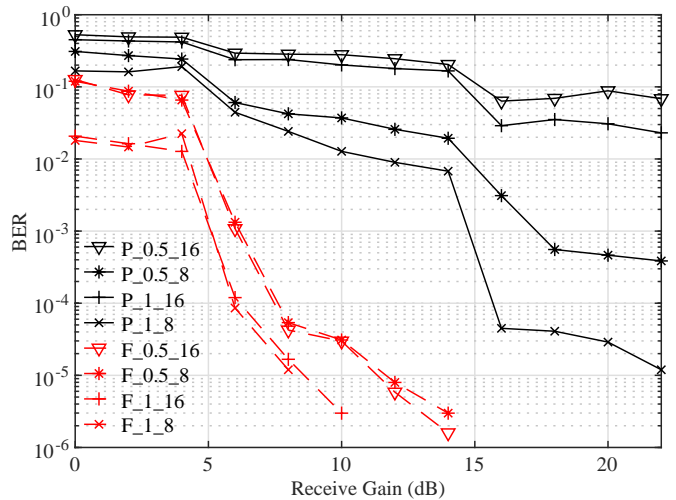}
	\caption{Curve of BER versus receive gain. Since the linearity of the actual gain curve is not very good, as shown in Fig. \ref{fig: Gain}, there is a certain error in the BER result of the actual test.} 
	\label{fig: SNR_BER}
\end{figure}

Fig. \ref{fig: SNR_BER} plots the BER versus SNR, where $800$ CPIs are
collected, corresponding to $1,024,000$ demodulation results. We see that the trends of all BER curves match well with what observed in Fig. \ref{fig: sim_SNR_BER}. This validates the proposed design in practical DFRC scenarios. 
The curves are not as smooth as the simulated ones, mainly because the gain control curve is not ideally linear and stable during the collection of four sets of data, as shown in Fig. \ref{fig: Gain}.

\section{Conclusion}  \label{sec: Conclusion}
In this work, a practical FH-MIMO DFRC is developed comprehensively treating all practically inevitable hardware errors, including STO, CFO and front-end imperfections of transceivers. We model these errors and analyze their impacts on FH-MIMO DFRC. Moreover, we design new waveforms and develop a low-complexity algorithm jointly estimating all hardware errors at a communication receiver. In addition, we build an FH-MIMO JRC experiment platform employing low-cost SDR and COST products that are popular in IoT system designs. 
Outdoor and indoor experiments are conducted using the platform. Applying the proposed designs on the collected experiment data achieves high performances for both radar and communications.

\appendix
\section{Illustrating $ T_{\mathrm{p}}\gg N_{\mathrm{p}}\dts $}     \label{app: Tp>> N Delta Ts}
As a radar PRT, $ T_{\mathrm{p}}=N_{\mathrm{p}}T_{\mathrm{s}}^{\mathrm{t}} $, where $ T_{\mathrm{s}}^{\mathrm{t}} $ denotes the sampling time at the radar transmitter. Thus, to compare $ T_{\mathrm{p}} $ and $ N_{\mathrm{p}} \Delta T_{\mathrm{s}}  $, we need to show that $ \Delta T_{\mathrm{s}}\ll T_{\mathrm{s}}^{\mathrm{t}} $. 

Since $ \dts $ is the sampling time difference between the radar transmitter and the communication receiver, we can perform the following calculation,
\begin{equation}  \label{eq: Ts}
	\begin{split} 
		\Delta {T_{\ms}}  = T_{\ms}^{\mt} -T_{\ms}^{\mr}
		= \frac{{f_{\ms}^{\mr} - f_{\ms}^{\mt}}}{{f_{\ms}^{\mt}f_{\ms}^{\mr}}} 
		= \frac{{ - \Delta {f_{\ms}}}}{    {{f_{\ms}}^{\mt}  ({f_{\ms}}^{\mt}  - \Delta {f_{\ms}})} } 
		= \frac{-\rho}{f_{\ms}^{\mt}(1-\rho)},
	\end{split}
\end{equation}
where we use the superscripts $(\cdot)^{\mathrm{t}}$ and $(\cdot)^{\mathrm{r}}$ to differentiate the variables between the transmitter and receiver, respectively; $ f_{\mathrm{s}}=1/T_{\mathrm{s}} $ denotes the sampling frequency; $ \dfs =f_{\mathrm{s}}^{\mathrm{t}}-f_{\mathrm{s}}^{\mathrm{r}} $; and $ \rho=\dfs/f_{\mathrm{s}}^{\mathrm{t}} $.
The reason of introducing $ \rho $ is that it can be linked with the clock stability at the radar transmitter. Specifically, we have
\begin{align}
	\rho = \dfs/f_{\mathrm{s}}^{\mathrm{t}}  = \dfclk/f_{\mathrm{CLK}}^{\mathrm{t}}, 
\end{align}
where $ f_{\mathrm{CLK}}^{\mathrm{t}} $ is the transmitter clock frequency, $ f_{\mathrm{s}}^{\mathrm{t}} $ is a multiple of 
$ f_{\mathrm{CLK}}^{\mathrm{t}} $, and $ \dfs $ is the same multiple of $ \dfclk $. The typical value of $ \rho $ is about tens of parts per million (ppm). Substituting the value into (\ref{eq: Ts}), we obtain $ \dts \approx 10^{-4}T_{\mathrm{s}}^{\mathrm{t}} \ll T_{\mathrm{s}}^{\mathrm{t}} $.

\bibliographystyle{elsarticle-num}
\bibliography{IEEEabrv,./ref/bib_JCAS.bib}
~~~\\
~~~\\

\end{document}